\documentclass[12pt]{iopart}

\usepackage{graphicx}

\begin{document}

\topical[What is so special about strangeness in hot matter?]{What is so
  special about strangeness in hot matter?%
\protect\footnote[1]{based on a student lecture given at the
    International Conference on Strangeness in Quark Matter, Atlantic
    Beach, NC, USA, March 12--17, 2003}}

\author{J\"urgen Schaffner--Bielich
\footnote[2]{e-mail: schaffner@th.physik.uni-frankfurt.de}
}

\address{Institut f\"ur Theoretische Physik,
J. W. Goethe Universit\"at,
D--60054 Frankfurt am Main, Germany}

\begin{abstract} 
  The production of strange particles in a hot medium as produced in
  collisions of heavy ions is considered one of the most important
  signals for the phase transition to a quark-gluon plasma. In
  the first part of this lecture, the theoretical description of
  strangeness production in hot matter is outlined for a gas of quarks
  and gluons and for a hadronic gas and its impact on the deconfinement
  phase transition. Then in the second part, constraints from the
  underlying chiral symmetry of Quantum Chromodynamics (QCD) are
  utilized to extract signals with strangeness for the chiral phase
  transition in hot matter.
\end{abstract}

\section{Introduction}

This review intends to introduce to non-experts the basic ideas about
the production of strange particles in hot, strongly interacting matter.
I focus on matter at high temperatures and low baryon density, i.e.\ 
setting the baryochemical potential to zero. These conditions are most
likely realized in the central region of the collision of heavy ions at
relativistic bombarding energies where strongly interacting QCD matter
can be probed in the laboratory. QCD exhibits a phase transition at high
temperatures, around $T_c\approx 170$ MeV, as seen numerically on
lattice gauge calculations (see e.g.\ \cite{Karsch04}). Hadrons, mesons
and baryons, are composite particles of quarks (and gluons) are
present at low temperatures. Quarks and gluons are confined within the
hadrons. Above the critical temperature, quarks and gluons are
(asymptotically) free and not bound to hadrons anymore, i.e.\ a quark-gluon
plasma (QGP) of deconfined quarks and gluons is formed.

The phase transition from a quark-gluon plasma to hadronic matter has
happened during the early universe, about $10^{-4}$ s after the
big-bang. Then the nucleons were formed during the deconfinement
transition. In terrestrial laboratories one explores this phase
transition by bombarding heavy nuclei at high energies and hunts for
signals of the formation of the quark-gluon plasma (for a recent
overview about the physics of the quark-gluon plasma, see
\cite{DirkReview}). 

Particles with strangeness have been considered to be a particular useful probe
of the quark-gluon plasma. There is a series of meetings dedicated to the topic of 
strange quarks in matter whose proceedings give an excellent overview of this field of 
research, see \cite{SQM91,SQM94,SQM95,SQM96,SQM97,SQM98,SQM00,SQM01,SQM03}.

The enhanced production of strange particles was predicted to be a
signal for the formation of a plasma of quarks and gluons in heavy-ion
collisions (for a review on the physics of strangeness production in
heavy-ion collisions see \cite{Greiner02}). Indeed, recently,
indications for the formation of a new form of strongly interacting
matter, the strongly interacting quark-gluon plasma, in
ultrarelativistic heavy-ion collisions have accumulated and
strengthened so that the discovery of the quark-gluon plasma has been
put forward by several of the most influential theoreticians in the
field (see \cite{QGP} for a list of review articles and
\cite{GML04,Mueller04,Shuryak04,Horst04,Wang04}).

I will not touch the physics of strange matter at finite baryon
density, in particular not the physics of the cold, dense quark matter
with strangeness, its phenomenon of color superconductivity and the
physics of strangeness in astrophysics.  Here, I refer the interested reader to the
extensive review articles about strange matter \cite{GS99o}, about
cold quark matter \cite{MarkReview,TomReview} and about strangeness in
astrophysics \cite{Madsen99o}.

The paper is organized as follows: first, I will discuss the
production of strange particles in a quark-gluon plasma, then its
production in a hadronic gas at finite temperature. The focus in that
section will be about the confined (hadron gas) and deconfined
(quark-gluon plasma) phase.  Then, I will address the issue of
strangeness production in terms of symmetries of QCD, i.e.\ the
chirally broken phase (hadrons) and the chirally restored phase
(quarks and gluons). Both descriptions should be mutually compatible
with each other, as one knows from lattice gauge simulations that both
transitions, the deconfinement and the chiral phase transition, happen
at the same critical temperature. Each section closes with a short
discussion of recent developments in the corresponding research fields.

\section{Strangeness and the deconfinement phase transition}

In the following sections, I discuss the production mechanisms for
producing strange particles for two distinctly different pictures:
First, in the deconfined state of free quarks and gluons, and second,
for a free hadron gas. Corrections due to interactions and dynamical
effects are shortly addressed at the end of the corresponding
subsections.

\subsection{Strangeness in a quark-gluon plasma}

In 1982, Rafelski and M\"uller demonstrated, that the production of
strange quarks will be enhanced in heavy-ion collisions, if a plasma
state of quarks and gluons is formed \cite{RM82}. The arguments were basically
twofold. First, the production threshold for the associated production
of strangeness via a pair of strange-antistrange quark pairs is
considerably smaller than the one for hadrons. Second, the
equilibration timescale for producing strange particles in a
quark-gluon plasma is much smaller than the one for a hadronic gas, so
that the produced strange particles are not suppressed by dynamical
effects and the corresponding number density is close to the equilibrium value.

The first argument is fairly easy to see. Consider the energy needed
to produce strange particles for a gas of quarks and gluons in
comparison to the one for a hadron gas. The associated production of a
strange-antistrange quark pair can proceed by the fusion of two gluons
or two (massless) light quarks, 
$$ q + \bar q \leftrightarrow s + \bar s \quad (q=u,d) \qquad g + g
\leftrightarrow s + \bar s $$ so that only an mass excess of
$$ Q_{qgp}= 2m_s\approx 200 \mbox{ MeV} $$ is involved. On the other
hand, hadronic strangeness production proceeds in free space via
NN$\to$N$\Lambda$K with a considerably larger mass difference of the
incoming and outgoing hadrons (the Q-value) of
$$ Q_{hg}=m_\Lambda + m_K - m_N \approx 670 \mbox{ MeV} \quad . $$
Hence, strangeness production should be considerably enhanced in a
quark-gluon plasma relative to that of a free hadron gas.

\begin{figure}
\centerline{
\includegraphics[width=0.2\textwidth]{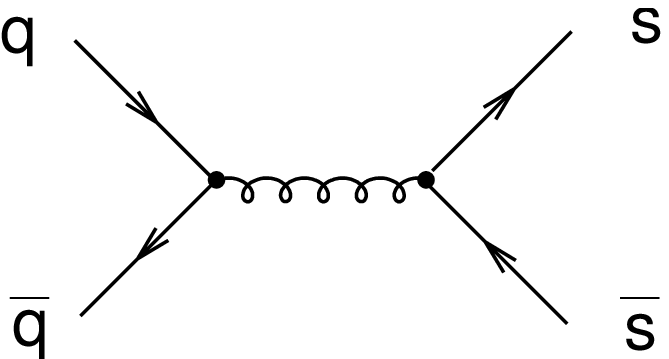}
\qquad
\includegraphics[width=0.2\textwidth]{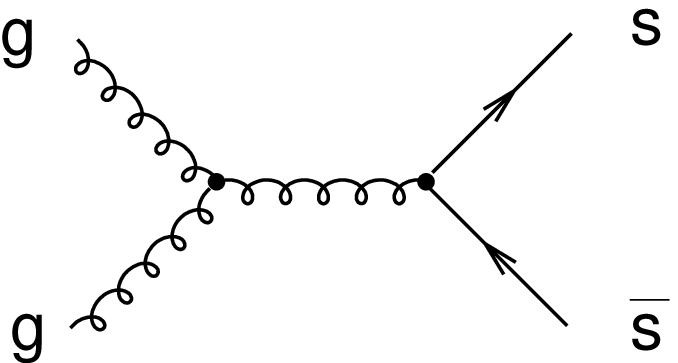}
\qquad
\includegraphics[width=0.2\textwidth]{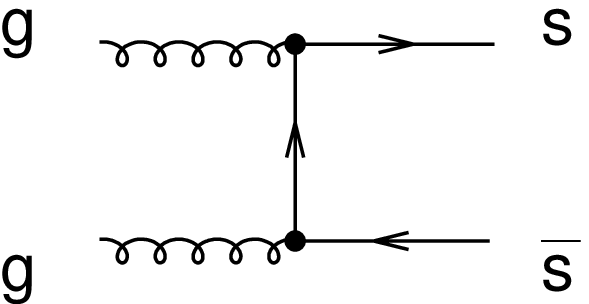}
\qquad
\includegraphics[width=0.2\textwidth]{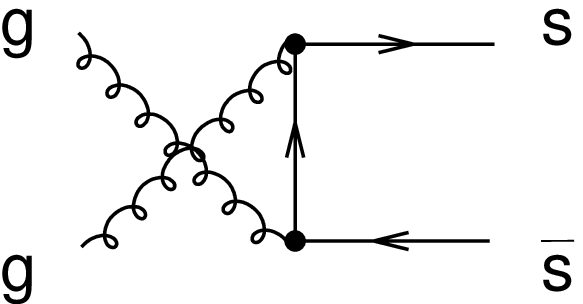}
}
\caption{The Feynman diagrams in perturbative QCD for the production of
  strange and anti-strange quarks in a quark-gluon plasma.}
\label{fig:feynman}
\end{figure}

\begin{figure}
\centerline{\includegraphics[width=0.6\textwidth]{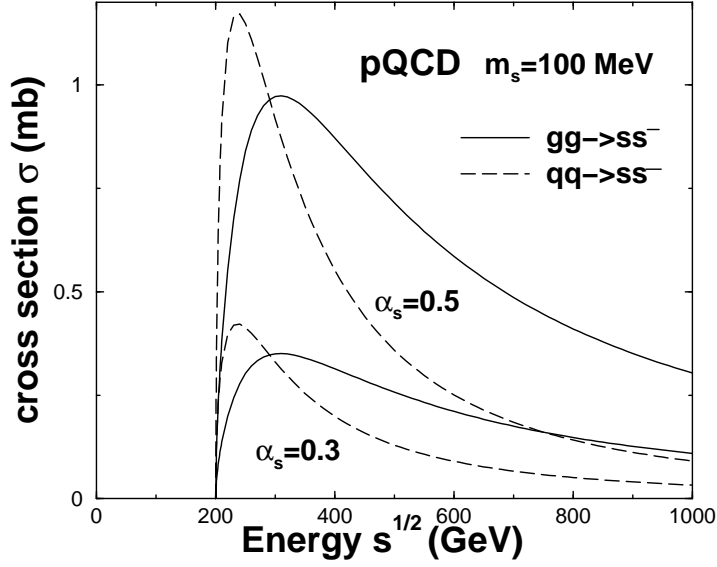}}
\caption{The cross section for strange-antistrange quark pair production
  via gluons (solid lines) and quarks (dashed lines) in perturbative QCD
  for a strange quark mass of $m_s=100$ MeV; upper lines are for
  $\alpha_s=0.5$, lower lines for $\alpha_s=0.3$.}
\label{fig:crosssection}
\end{figure}

For the second argument, a more elaborate and detailed calculation has
to be considered (see \cite{Field,Combridge79} for details). The Feynman
diagrams for the production of a strange-antistrange quark pair are
shown in Fig.~\ref{fig:feynman} to first order in perturbation theory.
Note, that there are three diagrams for the production process via gluons, where
one is due to the nonabelian character of the gluons in QCD.
Calculation of the Feynman diagrams gives for the cross section
involving quarks:
\begin{equation}
\sigma_{q\bar q\to s\bar s} = 
\frac{8\pi\alpha_s^2}{27s} 
\left( 1+\frac{2m_s^2}{s} \right)
\left( 1-\frac{4m_s^2}{s} \right)^{1/2}
= \frac{8\pi\alpha_s^2}{27s^2} \left(s+2m_s^2\right) \chi
\end{equation}
with 
$$
\chi = \sqrt{1-\frac{4m^2}{s}} \quad . $$
For $\alpha_s= 1/2$
and a typical energy scale of $s=(3T)^2\approx (0.6 \mbox{GeV})^2$ in a
thermal bath of massless particles, one finds a cross section of about
0.25 mb for $m_s=100$ MeV.  The gluonic production
processes result in a cross section of
\begin{equation}
\sigma_{gg} = \frac{\pi\alpha_s^2}{3s} \left[ \left(
1+\frac{4m_s^2}{s} + \frac{m_s^4}{s^2} \right) \mbox{ln} \left(
\frac{1+\chi}{1-\chi} \right)
-\left(\frac{7}{4}+\frac{31}{4} \frac{m_s^2}{s} \right) \chi \right]
\quad ,
\end{equation}
which turns out to be 0.6 mb for the same parameters. Hence, gluon
fusion is the dominant process for strangeness production in a
quark-gluon plasma. The cross section as a function of energy is
plotted in fig.~\ref{fig:crosssection} for two different values of the
strong coupling constant, $\alpha_s=0.5$ (upper lines) and
$\alpha_s=0.3$ (lower lines). The cross section has a threshold at
$\sqrt{s}=2m_s$, rises drastically, reaches a maximum just above the
threshold and falls down rapidly. 
The gluon production cross section for strange-antistrange
quarks dominates over the quark production cross section only for
larger energies well beyond the maximum.
Note, that the values of the
perturbative cross sections are quite small, in the range of 1 mb and
below, compared to a typical value of 40 mb for proton-proton
collisions. 

For the equilibration timescale, one has to look at the rate per unit
time and volume in a heat bath of given temperature. The cross section
has to be averaged over the distribution functions of the incoming
particles:
\begin{eqnarray}
A &=& \frac{dN}{dt\,d^3x} =
\frac{1}{2} \int_{4m_s^2}^\infty ds \cdot s \cdot \delta
\left(s-(k_1+k_2)^2\right)^2 \\[1ex] 
&\times& \int \frac{d^3k_1}{(2\pi)^3|k_1|} \int \frac{d^3k_2}{(2\pi)^3|k_2|}
\left\{ \frac{1}{2} (2\times 8)^2 f_g(k_1) f_g(k_2) \sigma_g(s) \right. \\[1ex]
&&\left. + 2\times (2\times 3)^2 f_q(k_1) f_{\bar q}(k_2) \sigma_q(s) \right\}
\end{eqnarray}
which corresponds to the analogue of the thermal average of the cross
section $<\sigma \cdot v>$ in the nonrelativistic case. Here, $f_g$ and
$f_q$ are the thermal distribution functions for gluons and quarks, 
respectively. Note, that no chemical potential is taken into account, the
calculation assumes that there is zero baryon density in the hot medium.
Also, Pauli-blocking effects are ignored. They turn out to be small, as
the phase space density of the produced strange quarks is not
sufficiently high as to Pauli-block the reactions.  Now, the density of
produced strange quarks in the hot medium as a function of time can be
derived by using the following master equation:
\begin{equation}
\frac{dn_s}{dt} = A \cdot \left\{ 1- \left(\frac{n_s(t)}{n_s(t=\infty)}\right)^2
\right\} 
\label{eq:nstime}
\end{equation}
where the strange quark density at infinite time $n_s(t=\infty)$
corresponds to the equilibrium density of strange quarks $n_s^{\rm eq}$.
The time evolution of the strange quark density eq.~(\ref{eq:nstime}) is
governed by a gain term and a loss term. The former one is just given by
the rate $A$, while the loss term is proportional to the squared density
of already produced strange quarks. The normalization is chosen in such
a way, that $n_s$ saturates for infinite times at the equilibrium value.
The equation can be formally solved for a constant rate to give
\begin{equation}
n_s(t) = n_s^{\rm eq} \cdot \mbox{tanh} \left( \frac{t}{\tau_{\rm eq}} \right) 
\end{equation}
where $\tau_{\rm eq}$ stands for the equilibration time scale as defined
by
\begin{equation}
\tau_{\rm eq} = \frac{1}{n_s^{\rm eq}\cdot A}
\quad .
\end{equation}
The time to reach a certain equilibrium fraction $f$ of strange quark
density relative to the equilibrium one is determined by
\begin{equation}
t_f = \tau_{\rm eq} \cdot {\rm tanh}^{-1}\left(\frac{n_s}{n^{\rm eq}_s}\right)
\end{equation}
This dependence of the equilibration time has to be compared to the
standard ones for an exponential behaviour of the number density (like
in ordinary radioactive decay) of $t_f = -ln(1-n/n_{\rm eq})$. For a
fraction of $f=1/e$, one gets $t_f=\tau_{\rm eq}$ and to reach an
equilibrium fraction of $f=0.99$ it takes $t_f = 4.6\tau_{rm eq}$ for
the exponential case. For our case of strange-antistrange quark
production in hot matter, which is proportional to the density {\em
  squared}, the times to reach a certain equilibrium fraction will be
shorter, i.e.\ $t_f \approx 0.74\tau$ for a fraction of $f=1/e$ and
$t_f\approx 2.6\tau$ for $f=0.99$. It is interesting to note, that a
similar physical system which obeys the characteristics discussed here
is the production of $^3$He in the proton-proton cycle in our sun whose
production rate depends also on the density squared (see pp.~340 in
\cite{Rolfs}).

To arrive at absolute numbers for the equilibration time scale, the
thermally averaged rate $A$ has to be calculated explicitly. For a
typical temperature of about $T=200$ MeV, the equilibration time turns
out to be $\tau_{eq}^{qgp}\approx 10$ fm, if a quark-gluon plasma is
formed \cite{RM82}. This equilibration time is about the timescale for
a relativistic heavy-ion collision from the initial collisions until
final freeze-out. Hence, it seems questionable that the system has
enough time to bring the production of strange quarks close to its
equilibrium value.

\begin{figure}
\centerline{\includegraphics[width=0.6\textwidth]{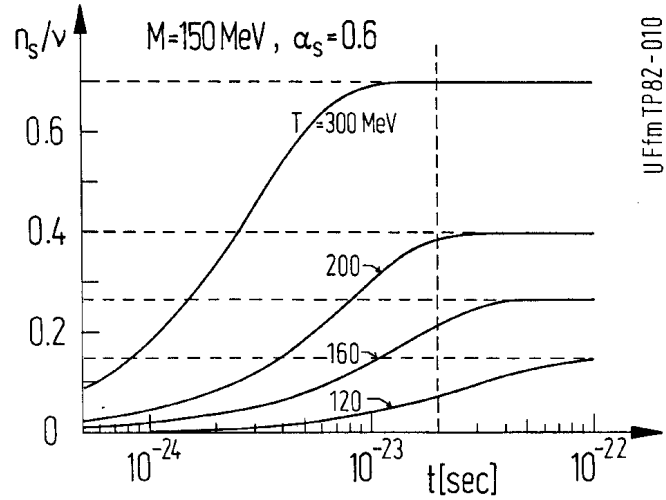}}
\caption{The number of strange quarks per baryon number in a hot
  gas of quarks and gluons as a function of the time. The solid
  horizontal lines denote the value in equilibrium for the different
  chosen temperatures. Reprinted figure with permission from
  \cite{RM82}. Copyright (1982) by the American Physical Society.}
\label{fig:nstime}
\end{figure}

Fig.~\ref{fig:nstime} shows the time evolution of the strange
quark density for different choices of the temperature in the plasma.
As one sees, the time to reach the equilibrium values depends strongly
on the temperature. For $T=300$ MeV, a time of about 3 fm seems to be
enough to fully equilibrate strangeness production in the plasma, while at a
temperature of $T=160$ MeV, something like 20 fm are needed. 

One should keep in mind, that it is assumed from the beginning, that
an equilibrated quark-gluon plasma is formed with a given
temperature. Of course, the quark-gluon plasma needs some time to be
formed, as well as the temperature will drop when the system
expands. Those dynamical effects are not taken into account in this
simple estimate.

\begin{figure}
\centerline{\includegraphics[width=0.6\textwidth]{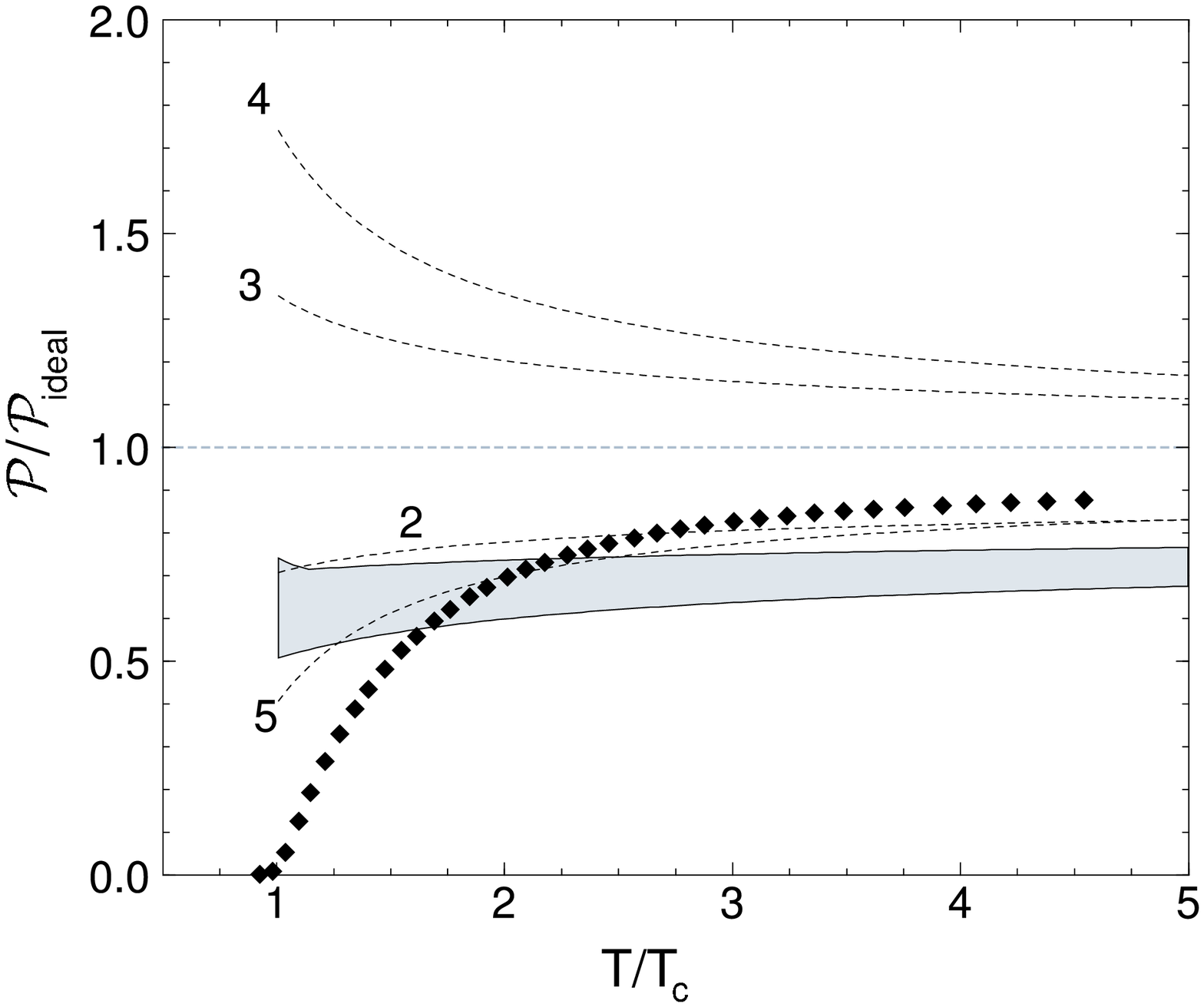}}
\caption{The contribution to the pressure at finite temperature for
  different orders of perturbation theory in QCD. One notes that the
  different orders oscillate indicating a break--down of perturbation
  theory even at temperatures well above $T_c$. Reprinted figure with
  permission from \cite{Andersen00}. Copyright (2000) by the American
  Physical Society.}
\label{fig:pqcdlattice}
\end{figure}

Moreover, which is even more severe, it is known from lattice data at
finite temperature, that perturbation theory fails and can not describe
the equation of state as extracted from the lattice. There are strong
nonperturbative effects even for temperatures which are up to 4 times
larger than the critical temperature, which has been measured to be
$T_c\approx 170$ MeV in full QCD (see e.g.\ \cite{Karsch04}).
Fig.~\ref{fig:pqcdlattice} depicts the various contributions in
perturbative QCD to the pressure as a function of temperature for each
order. While the second order calculation computes a pressure which is
below the one for an ideal gas of particles, the next order has a
different sign and brings the total pressure well above the ideal gas
pressure. Another change of sign occurs for the fifth order calculation
which even brings the pressure below the result for the second order
contribution. This oscillating behaviour and the fact that higher order
contributions are larger than the lower ones, demonstrates clearly, that
perturbation theory can not be used to describe the plasma of quarks and
gluons not even at the phase transition but also not up to $4T_c$!

Recent progress in finite temperature field theory utilizes resummation
techniques to tackle this problem with considerable more success than
pure perturbation theory like two--loop hard-thermal loop resummation
\cite{Andersen02,Andersen03}, for a review see \cite{Andersen04}.
Nonperturbative effects have been also included to some extent for the
calculation of the production of strange quarks, like the cut-off model
for gluons (see Fig.~11 in \cite{Rischke96}), massive gluons
\cite{Biro90}, and resummation by hard-thermal loops using different
approximation schemes \cite{Altherr93,Altherr94,Bilic95}. Massive gluons
allow for a new diagram of producing strange quarks, simply by the decay
of a massive gluon to a strange-antistrange quark pair. On the other
hand, a finite mass for gluons will suppress the gluon fusion processes.
An overall consistent picture of the nonperturbative effects in
equilibrium has not emerged yet, but the approaches studied so far
indicate that the equilibration time scale will stay above 10 fm when
nonperturbative effects are taken into account (see also the discussion
in \cite{Sollfrank95}). This picture is bolstered by the study of
equilibration time scales for quarks and gluons in \cite{Elliott99},
where it was found that equilibration in a quark-gluon plasma is
feasible for heavy-ion collisions at the future LHC but probably not at
RHIC energies of $\sqrt{s}=200$ GeV/nucleon.

There have been parton cascade models developed, which describe the
formation and the non-equilibrium expansion of the quarks and gluons
formed in a relativistic heavy-ion collision. The enhanced production of
strangeness has been studied using the VNI \cite{Geiger93} and the
HIJING model \cite{Csiz99}. Production of particles at large transverse
momenta, where methods of perturbative QCD are applicable, and impacts
for signaling the quark-gluon plasma are reviewed in detail by Gyulassy
\cite{Gyulassy04} taking into account nonperturbative effects (jet
quenching) from a strongly interacting quark-gluon plasma.

With the advent of the data from RHIC, at least two other new paradigms
of particle production substantiated: the notion of the formation of a
saturated state of gluons, the colour glass condensate
\cite{McLerran04}, and of quark coalescence \cite{Mueller04}. The latter
one is particular interesting for strange particles as they are more
sensitive to collective effects and the signals proposed by quark
coalescence. The relation of strange particle spectra and the colour
glass condensate has been partially explored via a generalized $m_t$
scaling behaviour in \cite{Schaffner01}.

It is clear from the above short discussion that strange
quark production in hot QCD is far from being settled and still is a very
active field of research.

\subsection{Strangeness in hadronic matter}

In this subsection, I discuss the production of strange particles in
the hadronic picture. The stable baryons in the vacuum under strong
interactions are besides the nucleons (N) the hyperons $\Lambda$ and
$\Sigma^{+,0,-}$ with one strange quark, the $\Xi^{0,-}$ with two
strange quarks, and the $\Omega^-$ with three strange quarks. The mass
of the baryons increases with the number of strange quarks. The stable
mesons under strong interactions are the pions $\pi^{+,0,-}$ and the
kaons $K^{+,0}$ with one anti-strange quark and its antiparticle
states $K^-$ and $\bar K^0$. Besides those stable particles, there are
more than hundred resonances known with a mass below 2 GeV which will
also appear in hot matter and will form resonance matter. For our
discussion, the first resonant state of the nucleon, the
$\Delta(1232)$ will be especially important in our following discussion. 

The production of strange particles in a free gas of hadrons has been
studied in Koch, M\"uller and Rafelski \cite{Koch86}. The production
of multiply strange baryons (the $\Xi$ and the $\Omega^-$) and of
antihyperons was found to be particularly strong suppressed in a
hadronic gas, as the equilibration timescales for their production was
much larger than the typical collision time of a heavy--ion collision.

\begin{figure}
\centerline{\includegraphics[width=\textwidth]{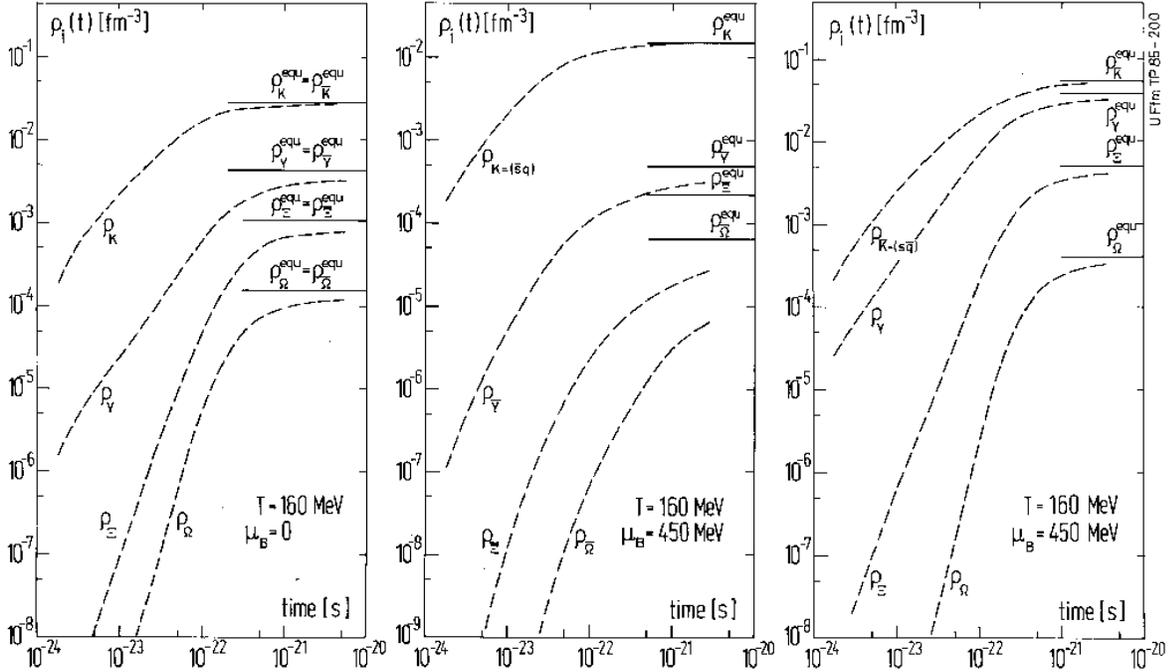}}
\caption{The number density of various strange hadrons for a temperature
  of $T=160$ MeV and different baryochemical potentials as a function of
  time. The solid horizontal lines denote the value in equilibrium
  for the different hadrons. Reprinted from "Strangeness in
  relativistic heavy ion collisions" \cite{Koch86}, Copyright
  (1986), with permission from Elsevier.}
\label{fig:eqtimehadron}
\end{figure}

The basic channels for strange hadron production in the vacuum are:
\begin{eqnarray}
\pi + \pi & \longrightarrow & K + \bar K \qquad
(Q= 2m_K - 2m_\pi \approx 710 \mbox{ MeV}) \\
N + N & \longrightarrow & N + \Lambda + K \qquad 
(Q = m_\Lambda + m_K - m_N \approx 670 \mbox{ MeV}) \\
\pi + N & \longrightarrow & K + \Lambda \qquad
(Q = m_\Lambda + m_K - m_N -m_\pi \approx 530 \mbox{ MeV})
\end{eqnarray}
where the latter one can be studied by a secondary beam of pions in
the laboratory.  The Q-values for these processes are $Q=710$ MeV, 670 MeV, and
530 MeV, respectively, so substantially higher than for a quark-gluon
plasma, where $Q=2m_s\approx 200$ MeV.  In the hot medium, as
stated above, resonances will appear so that channels like the
following are possible:
\begin{eqnarray}
N + \Delta & \longrightarrow & N + \Lambda + K \qquad 
(Q = m_\Lambda + m_K - m_\Delta \approx 380 \mbox{ MeV}) \\
\pi + \Delta &\longrightarrow& K + \Lambda \qquad 
(Q = m_\Lambda + m_K - m_\Delta -m_\pi \approx 240 \mbox{ MeV})
\end{eqnarray}
Now the Q-value is already comparable to the one for a quark-gluon
plasma! Even more, the Q-values can become smaller or even negative as for
\begin{eqnarray}
\Delta + \Delta &\longrightarrow& N + \Lambda + K \qquad \!\!\!
(Q = m_\Lambda + m_K + m_N - 2 m_\Delta  \approx 90 \mbox{ MeV}) \\
\pi + \rho & \longrightarrow & K + \bar K \qquad
(Q= 2m_K - m_\pi - m_\rho \approx 80 \mbox{ MeV}) \\
\rho + N & \longrightarrow & K + \Lambda \qquad
(Q = m_\Lambda + m_K - m_N -m_\rho \approx -100 \mbox{ MeV})
\end{eqnarray}
Of course, the abundances of resonances like the $\rho$ and the $\Delta$
are suppressed exponentially by their (higher) masses.  But the very low
Q-values achievable for reactions with resonances means that it is
possible that a resonance hadron gas might have production rates and
equilibration timescales which are close to the one for a quark-gluon
plasma. For a more quantitative answer one has to rely on a detailed
numerical computation which includes all known resonances and their
cross sections in hot matter.  Starting point is the evolution equation
for the change of each particle number
\begin{equation}
\frac{dN_i}{d^4x} = \sum_{j,k} <\sigma v>_i n_j(T) n_k(T) 
- \sum_{l} <\sigma v>_l n_i(T) n_l(T) 
\quad .
\end{equation}
The equation is a master equation consisting of a production term and
a loss term. Again, Pauli blocking effects are not taken into account,
as they are small corrections. The change of particle number is
proportional to the density of the incoming particles and the
thermally averaged cross section 
\begin{equation}
<\sigma v>\; \propto\; \int d^3p_1 d^3p_2 f_1(p_1) f_2(p_2) \sigma_{12} v_{12}
\end{equation}
where $f$ are the distribution functions of the particles, i.e.\
Fermi-Dirac or Bose-Einstein distribution functions for thermally
equilibrated matter. The cross section for most channels is poorly
known if known at all, like for $K + \Xi \to \Omega +\pi$, so that one
has to assume some universal cross sections for those cases. The
equilibration timescale one finds for a temperature of $T=160$ MeV are
shown in Fig.~\ref{fig:eqtimehadron} and can be read off to be about
$$ \tau_{\rm eq}^h \approx 10^{-22} \mbox{ s} \approx 30 \mbox{ fm} $$
for kaons, which is not so far from the one for a quark-gluon
plasma. But the timescale for the (anti)hyperons, especially the $\bar
\Xi$ and $\bar \Omega$ at finite density (finite baryochemical
potential), can be an order of magnitude longer! The reason is, that
it is more difficult to produce multiple units of strangeness in
hadronic processes than in a quark-gluon plasma, where the produced
strange quarks just coalesce to form a multiply strange baryon at
particle freeze-out.

Detailed follow-up calculations in transport models supported this
picture: the number of produced kaons in relativistic heavy-ion
collisions can be explained by a purely hadronic picture at AGS
energies \cite{Mattiello89} and at SPS energies \cite{Weber02}. Note,
that the often studied ratio of kaons to pions suffers from the fact
that their nonlinear behaviour as a function of bombarding energy
stems from the pions not the kaons \cite{Weber02}. Still, the kaon
slopes can not be explained in present transport models
\cite{Elena04}.

On the other hand, the production rates for antihyperons seems to be
underestimated by hadronic transport simulations. A number of possible
solutions have been proposed to remedy this, like colour ropes
\cite{Sorge92} or multiquark droplets \cite{Werner93}, which rely on the
quark-gluon picture.  More recently, it was demonstrated, that the
hadronic transport codes misses an essential reaction for a proper
description of antibaryon production: the annihilation process of two
antibaryons going to several mesons \cite{Rapp01,Greiner01}. The back
reaction, several mesons produce a pair of antibaryons, is not a
binary reaction, i.e.\ it involves more than two incoming particles,
which is not taken into account in present transport codes (for a most
recent attempt to incorporate $2\leftrightarrow 3$ consistently see
\cite{Greiner04}).  Nevertheless, the impacts of reactions like
$$
p + \bar p \leftrightarrow n \pi \qquad p + \bar\Lambda
\leftrightarrow K + n\pi 
$$ 
can be studied using master equations as introduced above. The updated
hadronic master equation models \cite{Rapp01,Greiner01} find now, that
the production rate of antibaryons will be also enhanced in a purely
hadronic approach by rescattering of multiple mesons into
baryon-antibaryon pairs. Equilibration times for (anti)hyperons to be
in chemical equilibrium can be as short as 10 fm as shown in
\cite{Kapusta02}.  Still, the production rates of the antihyperons
$\bar\Xi$ and $\bar\Omega$ can not be described in this scenario
unless effects from a phase transition and a rapid rise of the cross
section with the pion number density out of equilibrium are added to
the production rates
\cite{PBM03}.

An additional enhancement for strangeness production comes from the
fact, that the hadron masses will experience medium modifications. In
hot matter, the standard picture is that hadron masses decrease as a
function of temperature which will lower the Q-value of the
strangeness production processes even more and increase the number of
produced strange particles. It is well known that $\Lambda$ hyperons
feel an attractive potential in nuclei, thereby forming bound states
of a $\Lambda$ and a nucleus, a so called hypernucleus (see e.g.\
\cite{GS99o} and references therein). Also, the antikaons will be modified 
substantially which has been studied in a coupled channel calculation
for dense, cold matter \cite{Koch94,Waas96b,Lutz98,Ramos00,Tolos02} as
well as for hot matter \cite{SKE00,Tolos03}.

Strange particle ratios, for antikaons but also those involving
antihyperons, can be substantially enhanced in a hot medium in
particular close to a phase transition (see e.g.\
\cite{Scha91anti,Tolos03,Ziesche02} for calculations of particle
ratios including in-medium effects). I will address medium
modifications of hadrons in the next section utilizing chiral symmetry
of QCD to describe hadron masses at finite temperature and its
restoration as a measure of the phase-transition to a quark--gluon
plasma.

\section{Strangeness and Chiral Symmetry}

In the previous section, I have been looking at a free gas of quarks
and gluons in comparison to a free gas of hadrons. In this section, I
want to focus on a completely different approach to study signals of
the quark-gluon plasma with strange particles.

As pointed out earlier, it is known from lattice QCD simulations, that
there is a phase transition at a temperature of $T\approx 170$ MeV and
that there are highly nonperturbative effects even well above that
critical temperature. Lacking a detailed understanding of those
nonperturbative features of QCD in hot and dense matter, one has to
fall back on more fundamental features of QCD which do not need to
incorporate a detailed treatment of the interactions explicitly: symmetries. 

For the phase transition of pure gluon matter at finite temperature,
lattice QCD simulations demonstrate that the deconfinement phase
transition and the chiral phase transition coincide (see e.g.\ 
\cite{Kogut83,Kogut91}).
Note, that one can assign an order parameter for the deconfinement
phase transition, the Polyakov loop, only for the pure gluonic part of
QCD. Once quarks are included, only the chiral order parameter, the
quark condensate, remains as an order parameter which describes the chiral phase transition.
Hence, chiral symmetry plays a crucial role in describing the phase
transition of QCD in hot and dense, strongly interacting matter. Let
us be more specific now and take a look at the QCD Lagrangian for
three flavour massless quarks (q=u,d,s quarks):
\begin{equation}
{\cal L}^0_{qcd} = \bar q i \gamma_\mu \left(\partial^\mu-igA^\mu\right) q
- \frac{1}{2} \mbox{Tr} G_{\mu\nu} G^{\mu\nu}
\quad .
\end{equation}
The Lagrangian is invariant under the vector and axial transformations
\begin{equation}
q' = q + i \alpha^a \frac{\lambda^a}{2} q \qquad 
q' = q + i \beta^a \frac{\lambda^a}{2} \gamma_5 q
\end{equation}
with the corresponding conserved currents
\begin{equation}
V_\mu^a = \bar q \gamma_\mu \frac{\lambda^a}{2} q \qquad
A_\mu^a = \bar q \gamma_\mu \gamma_5 \frac{\lambda^a}{2} q
\quad .
\end{equation}
The Lagrangian exhibits a chiral SU(3)$_L \times $SU(3)$_R$
symmetry. One can assign now left-handed and right-handed quarks
\begin{equation}
q_L = \frac{1}{2} \left(1-\gamma_5\right) q \qquad
q_R = \frac{1}{2} \left(1+\gamma_5\right) q
\end{equation}
which transform separately and do not mix with each other.  Now in
reality, the quarks have a finite mass which adds the following terms
to the QCD Lagrangian:
\begin{equation}
\Delta {\cal L}_{mass} = - m_u \bar u u - m_d \bar d d - m_s \bar s s
\end{equation}
breaking the chiral symmetry explicitly. The mass terms mix the
left-handed and the right-handed quarks as
\begin{equation}
\bar qq = \bar q_L q_R + \bar q_R q_L
\quad . 
\end{equation}
Present estimates for the current quark masses range from
\begin{equation}
m_u = 1.5 - 5 \mbox{ MeV} \qquad
m_d = 3 - 9 \mbox{ MeV} \qquad
m_u = 60 - 170 \mbox{ MeV}
\end{equation}
as given by the Particle Data Group \cite{PDG02}. Note, that the up and
down quark masses are tiny in comparison to the nucleon mass
$m_{u,d}\ll m_N\approx 1$ GeV. The strange quark is much heavier than
the light quarks but still considerably lighter than the nucleon or
the hyperons, $m_s < 1$ GeV. The tiny masses of the light quarks are
actually essential to give the pion a finite mass. Nevertheless, a
reasonable and rather successful assumption is to describe QCD by
chiral symmetry plus corrections from explicit symmetry breaking.

Now, the masses of the hadrons are obviously not generated by quark
masses in our world. The major contribution to the hadron masses comes
from nonvanishing vacuum expectation values, i.e.\ from spontaneous
chiral symmetry breaking. The Gell-Mann--Oaks--Renner (GOR) relation
combines the nonvanishing expectation value for quarks, the quark
condensate, with the pion mass and the pion decay constant $f_\pi=92$ MeV:
\begin{equation}
m_\pi^2 f_\pi^2 = - \frac{1}{2} \left( m_u + m_d \right) < \bar uu + \bar dd >
\quad .
\end{equation}
The relation can be motivated heuristically: the right hand side stems
from the mass term of the QCD Lagrangian, the left hand side looks
like a mass term for the pion where the field it couples to has a
vacuum expectation value of just the pion decay constant. The equation
then connects the quark world to the hadron world where the lightest
known hadron is the pion. The GOR relation can be used to estimate the
value of the quark condensate which is
\begin{equation}
  <0|\bar qq|0> = - (310 \mbox{ MeV})^3 
\quad .
\end{equation}
assuming an average light quark mass of 5 MeV. Also the gluon fields
have a nonvanishing vacuum expectation value, the gluon condensate,
which according to QCD sum rules for charmonium states \cite{SVZ79}
amounts to
\begin{equation}
<0|\frac{\alpha_s}{\pi}G^a_{\mu\nu}G_a^{\mu\nu}|0> \approx (330 \mbox{ MeV})^4
\quad .
\end{equation}
At first glance, the quark and gluon condensate seems to be of equal
magnitude. However, the quark condensate has to be multiplied with the
current quark mass so that
\begin{equation}
<0|\frac{\alpha_s}{\pi}G^a_{\mu\nu}G_a^{\mu\nu}|0> 
\approx 80 <0|\bar qq|0> m_{u,d} 
\approx 4 <0|\bar qq|0> m_s 
\end{equation}
and the gluon condensate turns out to be much larger than the
corresponding contribution from the quark condensate. It is the gluon
condensate which generates the trace anomaly of QCD, i.e.\ a
nonvanishing vacuum expectation value for the trace of the
energy-momentum tensor
\begin{equation}
\theta_\mu^\mu = \frac{\beta_{\rm QCD}}{2g} G^a_{\mu\nu}G_a^{\mu\nu} 
+  \sum_{i=u,d,s} m_i \cdot \bar q_i q_i
\end{equation}
where $\beta_{\rm QCD}$ is the QCD $\beta$ function and $m_i$ the
current quark mass. The trace anomaly can be related to the vacuum
bag pressure \cite{SVZ79}, which basically determines the hadron masses in the MIT bag
model.

\begin{figure}
\centerline{\includegraphics[width=0.5\textwidth]{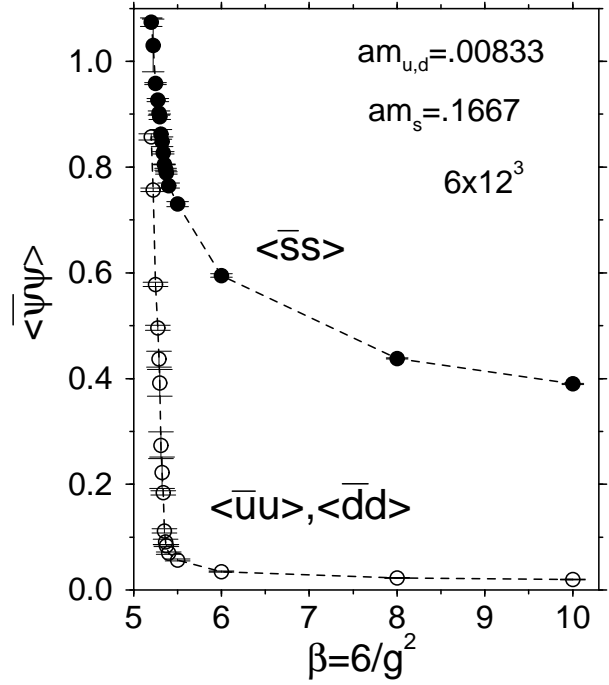}}
\caption{The lattice data for the light and strange quark condensate as
  a function of $\beta$ (a measure of the temperature). 
  The data is taken from \cite{Kogut91}.}
\label{fig:qqbarlattice}
\end{figure}

Now the quark and gluon condensates will change in a hadronic medium. In
particular at finite temperatures and zero density, the quark condensate
is an order parameter for the chiral phase transition (strictly only in
the chiral limit, i.e.\ for vanishing current quark masses).
The quark condensate will melt, its value will drop to zero at the
chiral phase transition temperature $T_\chi$. Lattice gauge simulations
demonstrated that the light quark condensate indeed decreases
drastically at a temperature of $T_c\approx 170$ MeV which coincides
with the deconfinement phase transition \cite{Karsch04}. Lattice data
also indicates that the strange quark condensate gets smaller at $T_c$,
although much less pronounced due to the larger current mass of the
strange quark \cite{Kogut91}. Fig.~\ref{fig:qqbarlattice} depicts the light
quark and strange quark condensate as a function of $\beta$ which is a
measure of the underlying temperature. One sees that the drastic drop of
the light quark condensate at large temperatures (larger values of
$\beta$) is accompanied by a moderate drop of the strange quark
condensate. Hence, also strange particles will be moderated by the
chiral or equivalently the deconfinement phase transition. Therefore,
strange hadrons can in principle be utilized as a signal for the chiral
phase transition. The advantage of using strange hadrons rather than
light nonstrange ones is that strange particles can carry information
from the high-density state as strangeness number is conserved in strong
interactions.

The main task is now to relate the behaviour of the quark condensates to
physical properties of (strange) hadrons. For that task one has to rely
on effective models which incorporate the symmetry constrains of QCD. In
the following, I will explore the effects of chiral symmetry
restoration for strange hadrons in a SU(3)$\times$SU(3) chiral model \cite{Gell60} at
finite temperatures. The hadrons involved are the pseudoscalar mesons
($\pi$,$K$,$\eta$,$\eta'$) and their chiral partners the scalar mesons
($\sigma$,$\kappa$,$a_0$,$f_0$) both forming a nonet in flavour
SU(3). All 18 mesons can be grouped in one complex matrix 
\begin{equation}
M=\Sigma + i\Pi = \sum_{a=0,8} \lambda_a \left(\sigma_a + i\pi_a\right)
\end{equation}
where $\lambda_a$ denote the Gell-Mann matrices.
One can form the following chiral invariants:
\begin{eqnarray}
\Tr M^\dagger M &\longrightarrow& \quad O(18)\qquad \mbox{\rm (norm of vector)} \\
\Tr M^\dagger M M^\dagger M &\longrightarrow& \quad U(3)\times U(3) 
\qquad (M\to U M U^{-1}) \\
\det M + \det M^\dagger &\longrightarrow& \quad SU(3)\times SU(3) 
\end{eqnarray}
where the right side denotes the corresponding symmetry of the
term. The last term breaks the $U_A(1)$ symmetry as $\det
\exp{(i\lambda_0)} =\exp{(i\Tr\lambda_0)} \neq 1$. The model exhibits
two order parameters, the expectation value of the $\sigma$ and the
$f_0$ field ($\zeta$) which can be associated with the light quark and
the strange quark condensate, respectively. In principle, there are
three Gell-Mann matrices which are diagonal and can be associated with
nonvanishing expectation values and order parameters, $\lambda_0$, $\lambda_3$ and
$\lambda_8$, but $\lambda_3$ only breaks isospin symmetry and can be
ignored in the following. The effective Lagrangian reads then
\begin{eqnarray}
{\cal L} &=& \frac{1}{2} \Tr \partial_\mu M^\dagger \partial^\mu M 
+ \frac{1}{2} \mu^2 \Tr M^\dagger M 
- \lambda \cdot \Tr \left(M^\dagger M M^\dagger M\right) \cr
&&
- \lambda' \cdot \left(\Tr M^\dagger M\right)^2
+ c \cdot \left(\det M + \det M^\dagger \right)
\end{eqnarray} 
which was used by Pisarski and Wilczek to study the order of the chiral 
phase transition in QCD \cite{Pisa84}.
The parameters of the model are $\mu^2$, $\lambda$, $\lambda'$, and
$c$. They are determined by a fit to $m_\pi$, $m_K$, $m_\eta^2 +
m_{\eta'}^2$ and $m_\sigma$. 
Explicit breaking of chiral symmetry is introduced in the model via
\begin{equation}
{\cal L}_{\rm esb} = \epsilon \cdot \sigma + \epsilon' \cdot \zeta
\end{equation}
which is fixed by the PCAC hypothesis generalized to SU(3)
\begin{equation}
\partial^\mu A_\mu^a = f_a m_a^2 \pi_a
\end{equation}
so that $<\sigma_0> = f_\pi = 92.4$ MeV and $<\zeta_0> = \sqrt{2} f_K -
f_\pi/sqrt{2} = 94.5$ MeV, where $f_\pi$ and $f_K$ are the pion and kaon
decay constants, respectively.

\begin{figure}
\centerline{\includegraphics[width=0.8\textwidth]{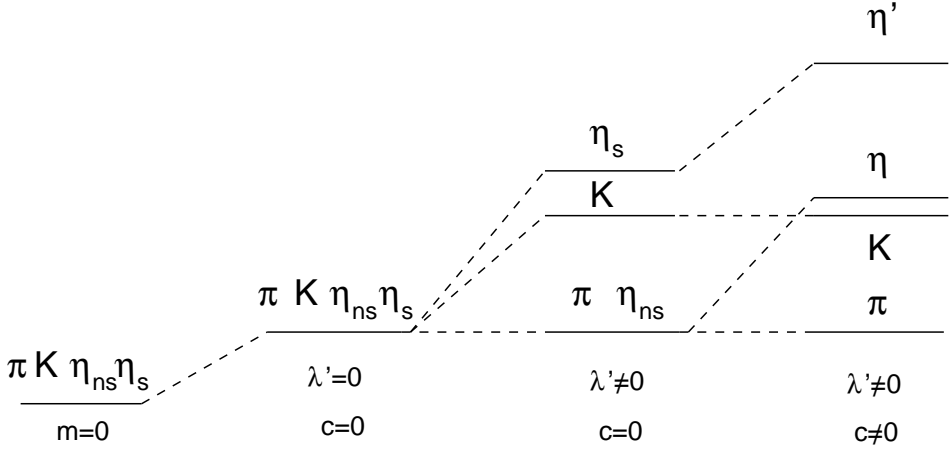}}
\caption{The mass splittings of the pseudoscalar mesons in the chiral SU(3)
  models.} 
\label{fig:mass_splittings}
\end{figure}

The mass splittings in SU(3) are characterized in
Fig.~\ref{fig:mass_splittings}. In the chiral limit, i.e.\ for vanishing
quark masses, there are nine Goldstone bosons and all nine pseudoscalar
mesons of the nonet are massless. The physical $\eta$ and $\eta'$ are
combinations of the singlet and octet state which mix in such a way that
they are equal to the nonstrange $\eta_{ns}$ and hidden strange
$\eta_{s}$ (this is the case of the so called ideal mixing).  
For a small but finite quark mass, there are
nine pseudo--Goldstone bosons which have a finite mass. For the case
that the coupling constants $\lambda'$ and $c$ are zero, all these
pseudo--Goldstone bosons are degenerate, even for different light and
strange quark masses. Letting $\lambda'\neq 0$ breaks the $O(18)$
symmetry and the masses split according to their quark content. The pion
and the $\eta_{ns}$ are degenerate in mass, the kaon is heavier and the
$\eta_{s}$ as a pure $\bar ss$--state is the heaviest one. In the last
column to the right, effects from the braking of the $U_A(1)$ anomaly are
switched on ($c\neq 0$). Now the $\eta'$ is a mixture of the $\eta_{ns}$
and $\eta_s$ and is the heaviest of all pseudoscalar mesons. The $\eta$
mass is slightly larger than the kaon mass. What happened during the
last stage is actually, that the degeneracy of the pion and the
$\eta_{ns}$ is lifted by the presence of the $U_A(1)$ anomaly and the
mass of the $\eta_{ns}$ is shifted even above the mass of the kaon and
$\eta_{s}$. As the two $\eta$ states mix with each other strongly, a
level crossing occurs in reality so that there is a continuous line in
mass shift between the $\eta_s$ and $\eta'$ and between the $\eta_{ns}$
and the physical $\eta$. Hence, the $\eta$ and not the $\eta'$ is
expected to have a large (hidden) strangeness content!

The considerations above are made for zero temperature but give a quite
accurate picture of what is going to happen at high temperatures when
chiral symmetry is restored. If $SU(2)\times SU(2)$ symmetry is
restored, the chiral partners $\pi$--$\sigma$ and $\eta_{ns}$--$a_0$ are
degenerate in mass separately:
\begin{equation}
m_\pi = m_\sigma  <  m_{\eta_{ns}} = m_{a_0}
\end{equation}
If in addition, the $U_A(1)$ anomaly is restored, these mass doublets
will be equal in mass:
\begin{equation}
m_\pi = m_\sigma = m_{\eta_{ns}} = m_{a_0}
\end{equation}
If there is at least a partial restoration of the
$U_A(1)$ symmetry at the critical temperature then
\begin{equation}
m_\pi = m_\sigma \approx m_{\eta_{ns}} = m_{a_0}
\end{equation}
The interesting and strange twist to this effect is, that the mass splitting
of the purely nonstrange meson doublets is governed by the strange quark condensate,
i.e.\ $\delta m^2 \propto c\cdot \zeta$!

That hadronic masses change in a hot medium has been known for quite
some time and confirmed by lattice gauge simulations (see e.g.\ 
\cite{Gottlieb97}). The pion mass rises as a function of temperature,
while the mass of the $a_0$ drops for temperatures below $T_c$. The
masses of the pion and the $a_0$ as well as for the pion and the kaon
are degenerate at high temperatures well above $T_c$. 
It seems that the
chiral anomaly is partially restored at $T_c$ as $m_{a_0}$ is close to 
$m_\pi$ at $T_c$  
(see \cite{Kuni89,Kuni91,Scha00l} and references therein).
It is interesting
to note, that the states are below the two--quark threshold so that the
quark-gluon plasma seems to support hadronic excitations
\cite{Gottlieb97}! Even more than that, there seems to be bound states
well beyond phase transition in the quark-gluon plasma at temperatures
of say $T=(1-3)T_c$ \cite{SZ04}. 

\begin{figure}
\centerline{\includegraphics[width=0.7\textwidth]{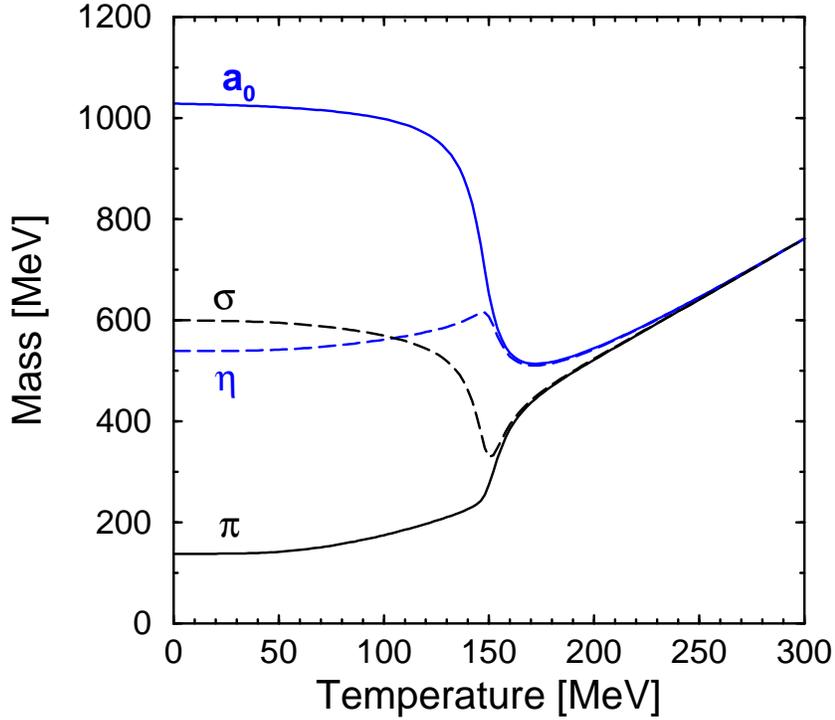}}
\caption{Meson masses as a function of temperature in an effective SU(3)
  chiral model. Note, that the mass for $\pi$ and $\sigma$ as well as
  for $a_0$ and $\eta$ are degenerate at $T_c\approx 150$ MeV and
  approach each other at larger temperatures (see \cite{Scha00l}).}
\label{fig:masses1}
\end{figure}

The pseudoscalar and scalar mesons masses can be studied in the SU(3)
linear sigma model at finite temperatures
\cite{Scha00l,Lena00}. Thermal fluctuations of the meson fields change
the order parameters $\sigma$ and $\zeta$ which will change the meson
masses. The whole set of equations for the thermal fluctuations, the
order parameters and the meson masses can be solved selfconsistently
at a given temperature. So far, the temperature dependence of the
chiral anomaly has been put in by hand, so that the $U_A(1)$ symmetry
is partially restored at $T_c$ and smoothly interpolates between the
limiting cases of fixed anomaly coefficient $c\neq 0$ and $c=0$. The
effects of the chiral anomaly on the meson masses have been also
studied in the Nambu--Jona-Lasinio model by Kunihiro
\cite{Kuni89,Kuni91} with qualitative similar results. 

Fig.~\ref{fig:masses1} shows the (pole) mass for the nonstrange mesons
$\pi$, $\sigma$, $\eta$ and $a_0$ as a function of temperature. At
small temperatures, the meson masses hardly change. Only for
temperatures close to $T\approx 150$ MeV, the masses of the scalar
mesons $\sigma$ and $a_0$ drop drastically. The pseudoscalar meson
masses increases slightly with temperature so that their curves merge
with their corresponding chiral partners at $T=150$ MeV. At this
temperature, chiral symmetry for the light quark sector is restored
(besides small effects from the explicit symmetry breaking). Still,
there is a substantial mass splitting between the chiral doublets of
$\pi-\sigma$ and $\eta-a_0$ due to the chiral anomaly which stems from
the (still) nonvanishing strange quark condensate! It is only for much
larger energies that all these nonstrange mesons become degenerate in
mass.

\begin{figure}
\centerline{\includegraphics[width=0.7\textwidth]{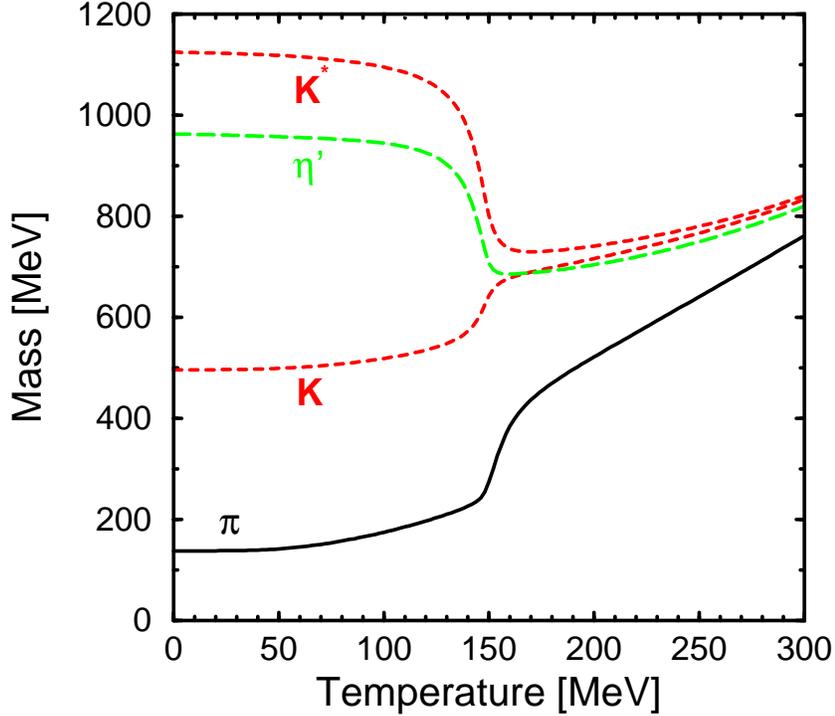}}
\caption{Meson masses as a function of temperature in an effective SU(3)
  chiral model. The masses for the scalar kaon $K^*$ and the pseudoscalar
  kaon $K$ approach each other at $T_c$ and above while the $\eta'$
  becomes even lighter than the kaon (see \cite{Scha00l}).}
\label{fig:masses2}
\end{figure}

Fig.~\ref{fig:masses2} shows the temperature dependence of the meson
masses of the kaon, $\kappa$ ($K^*$) and $\eta'$ in comparison to the
one for the pion. Like before, the mass of the scalar meson, the
$\kappa$, drops drastically, while the one for the kaon slightly
increase so that their curves approach each other at $T>150$
MeV. The mass of the $\eta'$ decreases with temperature, contrary to
the mass of the $\eta$, and even gets lower than the one for the
kaon. Note, that for temperatures larger than the crossover temperature
of $T=150$ MeV, the remaining mass splittings originates from the
strange quark condensate which melts less steeply at $T_c$ than the
light quark condensate. For large temperatures, $T\gg T_c$, all meson
masses will become degenerate.  

The drastic change of the mass spectra at finite temperatures should
have observable consequences for relativistic heavy-ion collisions.
Indeed, as the masses of the scalar mesons $\sigma$, $a_0$ and $\kappa$
fall so drastically close to the transition temperature, their decay
channels to pseudoscalar mesons, $\sigma\to\pi+\pi$, $a_0\to\eta+\pi$
and $\kappa\to K+\pi$, will be closed simply by phase space arguments!
The 'precursor' phenomenon of the chiral phase transition has been
pointed out in \cite{Hatsuda85} for the spectral function of the $\sigma$
meson, the analogue for the $a_0$ and $\kappa$ was studied in
\cite{Scha00l}.  While the measurement of $\eta$'s proceeds via leptonic
decays and is difficult to address experimentally, the decay products of
the $\sigma$ and $\kappa$ are readily measurable for a single collision
event. Note, that the number of combinations for the kaon plus pion mass
spectra is less than that for two pions as the kaon to pion ratio is
less than one (about 0.17 for central gold-gold collisions) making the
strange spectra more feasible experimentally.   

The invariant mass spectra of two pions was measured for heavy-ion
collisions of two gold nuclei at a bombarding energy of 200AGeV and used
to extract the $\rho$ meson \cite{Adams04} and the one for kaons plus
pions to get the $K^*(892)$ vector meson \cite{Adler02,Markert04}. So
far, the $\sigma$ meson as well as the $\kappa$ meson has not been
observed in these spectra. It will be difficult to extract clear signs
for the appearance of the $\sigma$ meson or the $\kappa$ meson, as the
resonances are only narrow around $T_c$ and the decay products can
rescatter and wash out any narrow resonance structure. On the other
hand, fast hadronization at $T_c$ will help to preserve the resonance in
the mass spectra of correlated pions and kaons. One might have to wait
for more complete and statistically precise data becoming available to
clarify this issue.

Nevertheless, it is clear that the masses of strange mesons will be
substantially reduced at the chiral phase transition, eventually being
close to the masses of nonstrange mesons. Hence, again the quark-gluon
plasma as being a chirally restored phase of strongly interacting
matter has the appealing feature that the production of strange
hadrons will be enhanced. Even more than that, there will be flavour
equilibration, i.e.\ nonstrange and strange particle yields are
similar to each other, as all masses are nearly degenerate above
$T_c$.

The questions remains how the chiral phase transition will influence
the mass spectra of other hadrons, in particular the masses of
(strange) vector mesons and baryons (hyperons) and its
antiparticles. This issue has been addressed in a chiral SU(3) model
at finite temperature and density \cite{Ziesche02}. The calculated
particle ratios change drastically when medium modified masses are
incorporated. The mass shifts are so strong that only temperatures and
densities below the chiral phase transition are compatible with the
experimentally measured particle ratios. In addition, a $\chi^2$ fit
to the particle ratio data demonstrates, that the freeze-out of
particles in relativistic heavy-ion collisions in equilibrium should
then just happen at a temperature of a few MeV below the chiral phase
transition temperature. As the mass changes so drastically around
$T_c$ it is unlikely that the system can adjust in the short timescale
of the collision and the particles have to emerge out of equilibrium
from the chirally restored phase.

\section{Summary}

In these lecture notes, the production of strange particles in a hot
strongly interacting medium have been described. Starting motivation for
these investigations was the idea that strange particles are more
abundantly produced in hot matter of quarks and gluons, the quark-gluon
plasma. The topic has been tackled from quite different points of view:
first from the fact the quark-gluon plasma constitutes a state of
deconfined matter, second from the observations from lattice QCD that
the phase transition at $T_c\approx 160$ MeV coincides with chiral
symmetry restoration, i.e.\ the quark-gluon plasma is the chirally
restored phase of QCD%
\footnote{ This is at least correct for large temperatures and zero net
  baryon density, it need not be the case for small temperatures and
  large densities, see the discussion in \cite{FPS01}.}.

In deconfined matter of quarks and gluons, the production rates for
producing strange-antistrange quark pairs should be larger compared to
that for strange-antistrange hadron pairs, as the threshold is
considerably smaller. In addition, the equilibration timescale for
strangeness production in quark-gluon plasma should be substantially
smaller than the timescale of a typical relativistic heavy-ion
collisions. Estimates based on an ideal gas of quarks and gluons
support the picture. However, lattice gauge simulations demonstrate
that there are strong corrections from the non-ideal behaviour of a
quark-gluon plasma even well above the phase transition
temperature. First attempts to incorporate nonperturbative effects
find that the equilibration timescale is shifted upwards close to the
dynamical scale of heavy-ion collisions.

The hadronic gas, for comparison, has to overcome a much larger Q-value
for the associated production of strangeness. This, however, holds
strictly only for binary collisions in free space. In a hot hadron gas,
scattering of secondary particles, resonances, lower the Q-value
drastically, so that light strange hadrons, as the kaons, can be
produced more easily. Antibaryons and in particular antihyperons can be
produced by multi-pion and kaon fusion processes. The arguments can be
extended to heavier hadrons with multiple units of strangeness, as the
$\bar\Xi$ or the $\Omega^-$, but seem to fail to describe the RHIC data
if not additional effects from a phase transition are
taken into account.

The medium effects of (strange) hadrons have been studied in a chiral
SU(3) model for the pseudoscalar and scalar mesons.  The hadron masses
in vacuum are basically determined by vacuum expectation values of the
quark and gluon condensates. For temperatures below $T_c$ the hadron
masses hardly change as a function of temperature. However, close to
$T_c$, the mass spectra dramatically changes. As chiral symmetry gets
restored, the quark and gluon condensates melt. Then, the masses of
the chiral partners of pseudoscalar and scalar mesons
($\pi$--$\sigma$, $\eta$--$a_0$, $K$--$\kappa$) have to become
degenerate. There is some intrinsic mass splittings between some
chiral partners due to the chiral $U_A(1)$ anomaly which is
proportional to the strange quark condensate.  The chiral anomaly
diminishes as the strange quark condensate vanishes.  The mass spectra
at temperatures above $T_c$ approaches a flavour independent one,
establishing flavour equilibration in the chirally restored phase.
Calculation of a hot medium taking into account effects from chiral
symmetry restoration finds that the hadron masses can not
substantially change in order to get the experimentally observed
particle ratios.  This might indicate, that the particles have to
freeze-out closely below the chiral phase transition.

It is interesting to see, how the two pictures of the phase transition
of QCD seem to merge into an overall consistent picture. However, I
think there needs to be a lot of work to be done and I hope this
article will stimulate further research to finally settle the issue of
strangeness production and the phase transition in hot and strongly
interacting QCD matter.

\ack

I thank 
Peter Braun--Munzinger, 
Adrian Dumitru,
Carsten Greiner, 
Peter Koch--Steinheimer, 
Berndt M\"uller, 
Dirk Rischke, 
Igor Shovkovy, 
Johanna Stachel, 
Horst St\"ocker,
Detlef Zschiesche, 
for helpful discussions during the course of this writing.
I am indebted in particular to Berndt M\"uller for his idea and
encouragement to write up this article from the notes of my student
lecture given at the Strange Quark Matter meeting in 2003.

\section*{References}

\bibliographystyle{prstym}
\bibliography{all,literat,conf}

\end{document}